\documentstyle[aps,prl,epsfig]{revtex}
\begin{document}
\twocolumn[\hsize\textwidth\columnwidth\hsize\csname@twocolumnfalse\endcsname

%\draft command makes pacs numbers print
\draft

\title{
Dynamics of a Tube Conveying Fluid
}

\author{
S. Shima\cite{mail} and T. Mizuguchi
}
\address{
Department of Physics, Graduate School of Science, 
 Kyoto University, Kyoto 606-8502, JAPAN
}
\date{\today}
\maketitle

\widetext

\begin{abstract}
A tube conveying a large amount of fluid with a free outlet does not
 sit still.
%, as anyone knows from empirical experience. 
We construct
 and analyze a nonlinear evolution equation describing such phenomena.
 Two types of boundary conditions at the inlet are considered, one
 for which 
 it is clamped and one for which it is hinged.
 Analyzing the linear stability of
 the trivial solution, we find that with the former boundary conditions,
 it exhibits a ``flutter'' instability, while with the latter boundary
 conditions, it exhibits a ``rotation'' instability.
 These instabilities and the nonlinear behaviors of the system are also studied numerically.
\end{abstract}

]

\narrowtext

% body of paper here
Thready structures appear in a wide variety of physical systems and
exhibit many different distinctive patterns and types of motion.
For example, the behavior of chain polymers is a commonly studied topic
 in physics and chemistry. 
In fluid physics, the structures and dynamics of vortex filaments 
have been studied in detail. 
%Biological systems provide us with many interesting 
Biological systems provide us with many 
phenomena involving thready objects 
from microscopic to macroscopic scales, e.g.,
the behavior of DNA, the folding of proteins, 
the beating of flagellum, and the locomotion of snakes.
These phenomena are observed in non-equilibrium systems 
in which several factors, such as elasticity, driving forces 
 and dissipation, are balanced. 
In this class of phenomena, the motion of a tube conveying fluid with a
free outlet 
has been extensively studied, owing to its simple nature.
Paidoussis theoretically showed that
% the trivial, completely straight state becomes unstable at some flow rate,
 the trivial, straight state becomes unstable at some flow rate,
 and the tube begins to flutter as the result of a Hopf bifurcation 
\cite{1}.
 This result has been confirmed by several theoretical and
 experimental studies
\cite{2,3,4,5,6}.
The theoretical models used in these studies
 are physically very realistic, but they are applicable only for small
 amplitude motion. Also, in most studies, only one type of
 boundary conditions (BC) is considered, that of the cantilevered type,
 in which the tube is clamped at the inlet and free at the outlet. 
 In this Letter, we adopt a rather phenomenological
 approach to construct the minimal model for
 motion of any size amplitude under more general BC at the inlet.
 We obtain an evolution
 law in the form of a nonlinear integro-differential equation.
%A linear stability analysis of the trivial, straight solution
A linear stability analysis of the trivial solution
% suggests us not only the existence of a ``flutter'' instability but also
 suggests not only the existence of a ``flutter'' instability but also
the existence of a ``rotation'' instability, depending on the BC at the inlet. 
 Numerical simulations confirm the existence of these instabilities
% and elucidate the nonlinear behaviors of the equation. 
 and elucidate the nonlinear behavior of the equation. 

We begin by deriving  the evolution equations of the system. 
For this purpose, we first clearly define the physical system under
 consideration.
 Consider a tube conveying fluid with a free outlet.
%For simplicity we consider the case in which the motion is
The motion is
% restricted to $x$-$y$ plane, and the gravitational force is ignored.
 restricted to $x$-$y$ plane, and the gravitational force is not considered.
 Also, the length of the tube is regarded as fixed.
We treat the tube as one-dimensional structure and ignore any dependence
on its radius.
% There is an elastic force acting on the tube in response to
 An elastic force acts on the tube in response to
its being bent from a straight shape.
 In the tube, fluid flows at a constant rate,
 and its momentum change creates a force acting transversely on the tube.
A resistive force representing some kind of frictional interaction with a
surrounding medium is also included. 
As typical cases, we consider two types of BC at the inlet.
 One case is that in which there is a clamp at the inlet, keeping both
 the position and direction fixed.
The other is that in which there is a hinge at the inlet, keeping only
 the position fixed.
\begin{center}
\begin{figure}
\epsfig{file=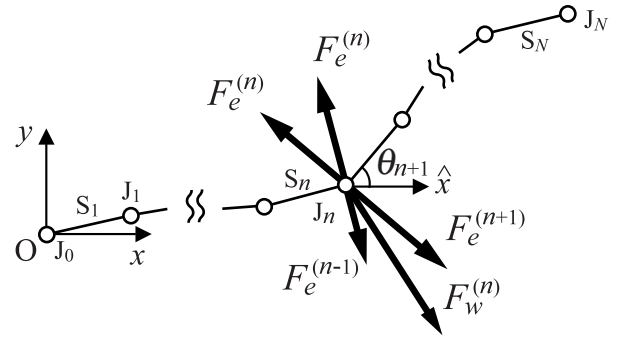}
 \caption{
%A discrete model of a tube consisting of $N+1$ joints and $N$
% segments. J$_n$ and S$_n$ denote the $n$th joint and segment, respectively.
A discrete model of a tube. 
J$_n$ and S$_n$ denote the $n$th joint and segment, respectively.
The $F_e^{(n-1)}$, two $F_e^{(n)}$, and $F_e^{(n+1)}$, all of the elastic
 forces that exert on J$_n$, act perpendicularly
 to the segments, and tend to straighten the tube.
% $F_w^{(n)}$, the transverse force resulting from fluid flow, divides
 $F_w^{(n)}$, the force resulting from fluid flow, divides
 the angle
$\protect\angle$J$_{n-1}$J$_n$J$_{n+1}$
 in half, and tends to increase the bending.
}
 \label{force}
\end{figure}
\end{center}

We assume that a tube with
 the properties described above can be obtained as the continuum limit
 of the discrete articulated model depicted in  Fig.\ref{force}.
 In this model, there are $N+1$ joints and $N$ segments.
 Let $(x_n,y_n)$ be the coordinates of the $n$th joint, and let $\theta _n$
 be the angle between the $n$th segment and the $x$ axis.
% Each joint has mass $m(N)=M/N$, where $M$ is the total mass of the
 Each joint has mass $m(N)\equiv M/N$, where $M$ is the total mass of the
 tube, and the segments have 
% no mass. The fixed length of each segment is $L/N$, where $L$ is
 no mass. The fixed length of each segment is $l(N) \equiv L/N$, where $L$ is
 the total length of the tube.
Each joint exerts an elastic force on itself and
 its neighboring joints that tends
 to straighten the tube.
 The magnitude of the force exerted is proportional to the difference
 between the angles of the segments on both sides:
 $F _e^{(n)} = k(N)\cdot |\theta _{n+1} -\theta _n|$
 for $1\le n \le N-1$, where $k(N)$ is a positive coefficient
 representing the stiffness of the tube. 
The momentum change of the fluid flow at a joint creates a
 transverse force, which acts to increase the bending. 
 The magnitude of this flow force is also proportional to the
 difference between the angles on both sides:
 $F_w ^{(n)}= w(N)\cdot |\theta _{n+1} -\theta _n|$
 for $1\le n \le N-1$
 where $w(N)$ is a positive coefficient representing the flow rate.
 Here, we assume that $|\theta _{n+1} -\theta _n|\ll 1$.
 The directions and the joints on which the forces 
 $\{\mbox{\boldmath $F$}_e^{(n)}\}$ and $\{\mbox{\boldmath
 $F$}_w^{(n)}\}$
 act are exhibited in Fig.\ref{force}. 
 In addition to the above described forces, resistive forces act on each joint in the form
 ${\mbox{\boldmath $F$}_v^{(n)}}=-c(N)\cdot\mbox{\boldmath $v$}_n $
 for $1\le n \le N$, where $\mbox{\boldmath $v$}_n$ is the velocity
 of $n$th joint and $c(N)$ is a positive coefficient representing 
the strength of the resistance.
 Since each segment has a fixed length,
 there are $N$ constraints. 
 We account for
 these constraints by using Lagrange multipliers.  
 As mentioned above, the inlet is either clamped or hinged.
 These conditions are realized by setting 
 $F_e^{(0)}=k\theta _1 $ if it is clamped
 and $F_e^{(0)}=0$
% if it is hinged, where
 if hinged, where
 $F_e^{(0)}$ is a force that acts only on the first joint J$_1$.   
% $F_e^{(0)}$ is a force acting only on the first joint J$_1$.   

Our next step is to derive evolution equations  for this discrete model.
To simplify the notation, we first restrict ourselves to the
% To simplify the notation, we first focus on the
 case of a clamped tube ($F_e^{(0)}=k\theta _1 $).
% case of a clamped tube, i.e., $F_e^{(0)}=k\theta _1 $.
 It is easy to write down $2N$ Newton's equations of motion containing $N$
% undetermined multipliers to describe the system.
 undetermined multipliers.
%The positions  $\{x_n,y_n\}$ in these equations can be rewritten in
The positions  $\{x_n,y_n\}$ can be rewritten in
 terms of the angles  $\{\theta _n\}$
 using the relation
% $(x_n,y_n)=(L/N)\sum_{i=1}^n(\cos{\theta_i},\sin{\theta_i})$.
 $(x_n,y_n)=l\sum_{i=1}^n(\cos{\theta_i},\sin{\theta_i})$.
 Eliminating the $N$ multipliers 
 from these equations yields $N$ independent evolution
 equations for the generalized coordinates $\{\theta _n\}$.
 Assuming that inertial terms, such as those containing
 $\ddot\theta_i$, $\dot\theta_i^2$, are negligible 
(Physically, this is justified for a strongly dissipative system.),   
 we obtain the rather simple equation
 \begin{equation}
 \label{discrete}
 C\sum _{j=1}^N{\cal F}_{ij}\dot \theta _j
 =\sum _{j=1}^NA\cdot {\cal G}_{ij}\theta _j
			    -B\cdot {\cal H}_{ij}(\theta _j-\theta _{j-1}). 
 \end{equation}
% Here, $A\equiv kN/(mL)$, $B\equiv wN/(mL)$,
 Here, $A\equiv k/(ml)$, $B\equiv w/(ml)$,
 $C\equiv c/m$, $\theta _0\equiv 0$, and
 ${\cal F}_{ij}$, ${\cal G}_{ij}$ and ${\cal H}_{ij}$ are the components of
 $N\times N$ matrices defined as
$
{\cal F}_{ij}={\cal F}_{ji}\equiv (N-j+1)\cos(\theta _i -\theta _j)
$ for $j\ge i$,
$
{\cal G}_{ij}={\cal G}_{ji}\equiv
      (-2+\delta _{iN})\delta _{ij}+\delta _{i,j+1}
$ for $j\ge i$, 
$
{\cal H}_{ij}\equiv\cos{\{(\theta _j +\theta _{j-1} -2\theta_i)/2\}}
$ for $j> i$,
$
{\cal H}_{ij}\equiv 0
$ for $j\le i$.
%We call a tube described by (\ref{discrete}) a non-inertial tube.
% The effect of inertial terms this equation does not include will
% be touched upon at the end of this Letter. 

 The final step in constructing our model is to take the continuum
 limit $N\to \infty$ of
 (\ref{discrete}). In this limit, the discrete index $n$ can be
 regarded as a continuous variable $s\in [0,L]$. 
Carrying out an integration by parts of the equation so obtained,
we can write
 the evolution equation for the continuous model as
 \begin{eqnarray}
 \label{continuum}
 &&\int_s^L{\rm d}s'\int_0^{s'}{\rm d}s''\left\{
 \gamma \partial _t\theta (s'')
	      \right\}
 \cos{\{\theta (s)-\theta (s'')\}}\nonumber\\
 &=&\alpha\partial _s^2\theta+\beta\sin{\{\theta(s)-\theta(L)\}}
 ,
 \end{eqnarray}
 with two BC:
 \begin{equation}
 \label{B.C.cl}
 \theta(0)=\partial _s\theta(L)=0.
 \end{equation}
 Here,
 $\alpha\equiv\lim_{N\to\infty}[l^4A(N)]\mbox{, }
\beta\equiv\lim_{N\to\infty}[l^2B(N)]$ and
$\gamma\equiv\lim_{N\to\infty}C(N)$.
These are constants independent of $N$.
% that are independent of $N$,
The convergence of these three limits 
  is assumed, so that the three terms 
  corresponding to the elasticity,
 the driving force resulting from
 fluid flow, and the dissipation are balanced.

 In the case of a hinged tube ($F_e^{(0)}=0$), the evolution equation
 is derived in the
  very same way, although the result differs slightly from that of a
  clamped tube.
  For the discrete model, in this case the above definition of
 ${\cal G}_{ij}$ is replaced with 
$
{\cal G}_{ij}={\cal G}_{ji}\equiv
      (-2+\delta _{iN}+\delta _{i1})\delta _{ij}+\delta _{i,j+1}
$ for $j\ge i$.
 Then, for the continuous model, the two BC are not
 those given by (\ref{B.C.cl}), but rather,
 \begin{equation}
 \label{B.C.hi}
 \partial _s\theta (0)=\partial _s\theta (L)=0.
 \end{equation}
We see below that this slight difference has a great influence on
  the behavior of the whole system.

 Note that we can always set $\gamma /\alpha=1$ and $L=1$ in
 (\ref{continuum}) through rescaling of
 $t$ and $s$. This means that the number of essential parameters governing
 the behavior of the non-inertial tube is only one, say,
 $\beta /\alpha \equiv \epsilon >0$.
  Note that an increase of the flow rate corresponds to an 
 increase of $\epsilon$. This rescaling leaves (\ref{continuum}) as
 \begin{eqnarray}
 \label{rescaled}
 &&\int_s^1{\rm d}s'\int_0^{s'}{\rm d}s''\left\{
 \partial _t\theta (s'')
	      \right\}
 \cos{\{\theta (s)-\theta (s'')\}}\nonumber\\
 &=&\partial _s^2\theta+\epsilon\sin{\{\theta(s)-\theta(1)\}}
 .
 \end{eqnarray}

We have thus derived the evolution equations for a clamped tube as 
 (\ref{rescaled}) with (\ref{B.C.cl}) and for a hinged tube as
 (\ref{rescaled}) with (\ref{B.C.hi}). In each case we set $L=1$.
Both systems have the same, trivial
 solution, $\theta(s)=0$, which corresponds to a completely straight tube.
We now investigate situations in which this trivial
  solution becomes unstable by considering the linearization of  
 (\ref{rescaled}) about this solution.

We first consider the case of a clamped tube. Assuming that for all
  $s\in [0,1]$, $|\theta(s)-0|\ll 1$ holds, we omit
all but linear order terms in (\ref{rescaled}) to obtain
 \begin{equation}
 \label{intlinear}
 \int_s^1{\rm d}s'\int_0^{s'}{\rm d}s''
 \partial _t\theta (s'')
 =\partial _s^2\theta+\epsilon\{\theta(s)-\theta(1)\}
 .
 \end{equation}
Unlike  the nonlinear equation (\ref{rescaled}),
% the integrand in 
% (\ref{intlinear}) depends only on $s''$. Thus, we can eliminate
 the integrand depends only on $s''$. Thus, we can eliminate
 the integrals by differentiating (\ref{intlinear}) twice. Here,
in order to retain all the information contained in (\ref{intlinear}),
two conditions must be added to the differential equation that results
 from this procedure: Representing (\ref{intlinear}) as
 $F(s)\equiv [(\mbox{l.h.s.}-\mbox{r.h.s.})\mbox{ of (\ref{intlinear})}]
 =0$, this equation is equivalent to the differential equation
 $\partial_s^2F(s)=0$, together with the conditions $\partial_sF(0)=F(1)=0.$
We thus find that (\ref{intlinear}) is equivalent to
 \begin{equation}
 \label{linear}
 \partial_t\theta=-\partial _s^4\theta-\epsilon\partial_s^2\theta,
 \end{equation}
 with the two extra BC
 \begin{equation}
 \label{B.C.lin}
 \partial_s^3\theta(0)+\epsilon\partial_s\theta(0)=\partial_s^2\theta(1)=0.
 \end{equation}

 Now, our objective is to examine the eigenfunctions of (\ref{linear})
 with the four BC given in (\ref{B.C.cl}) and
 (\ref{B.C.lin}). Expecting the existence of complex eigenvalues, we regard
 $\theta(s,t)$ as a complex variable $\Theta(s,t)\in
 \mbox{\boldmath $C$}$. Here,
 the linearized equation  we consider is
 $
 \partial_t\Theta=-\partial _s^4\Theta-\epsilon\partial_s^2\Theta,
 $
 with the four BC
 $
 \Theta(0)=\partial_s^3\Theta(0)+\epsilon\partial_s\Theta(0)=
 \partial_s\Theta(1)=\partial_s^2\Theta(1)=0
 $.
 Assuming the form 
 $\Theta(s,t)\equiv e^{(\sigma+i\omega)t}\Phi(s)$, with
 $\sigma,\omega\in\mbox{\boldmath $R$}$, $\Phi\in 
 \mbox{\boldmath $C$}$, we get
 \begin{eqnarray}
 \label{phi}
 &&\qquad\qquad\quad-\Phi''''-\epsilon\Phi''=(\sigma+i\omega)\Phi,\\
 \label{B.C.phi_head}
 &&\quad\Phi(0)=\Phi'''(0)+\epsilon\Phi'(0)=0\quad\;\;\;\mbox{clamped inlet,}\\
 \label{B.C.phi_tail}
 &&\quad\Phi'(1)=\Phi''(1)=0\quad\qquad\qquad\;\:\,\mbox{free outlet.}
 \end{eqnarray}
For a given $\epsilon$, the set of all such solutions
 $\{(\sigma,\omega,\Phi(s))\}$ may span the solution space of
 (\ref{phi}) with BC given in (\ref{B.C.phi_head}) and (\ref{B.C.phi_tail}).
However, all we want to know at present is
  when and how the trivial solution becomes unstable.
For this reason, we set $\sigma=0$ in (\ref{phi}) to obtain
 \begin{equation}
 \label{phi2}
 -\Phi''''-\epsilon\Phi''=i\omega\Phi.
 \end{equation}

We now proceed to examine the solution set of (\ref{phi2}),
with the BC given in (\ref{B.C.phi_head}) and (\ref{B.C.phi_tail}).
First, we assume that the trivial solution becomes unstable beyond
 some critical value 
$\epsilon_{cr}$. Then, there necessarily exists
 at least one solution,
$(\epsilon_{cr},\omega_{cr},\Phi_{cr})$.
Note that this $(\epsilon_{cr},\omega_{cr},\Phi_{cr})$ need not be
the critical mode itself when the $\sigma =0$ eigenspace is non-simple.
In fact, such case exists for the hinged system.
 
The general solution of (\ref{phi2}) is easily obtained.
The characteristic polynomial of this equation is 
$
L(P)\equiv P^4+\epsilon P^2+i\omega=0.
$
Let us denote the four roots of $L(P)=0$ as
 $\pm P_1(\epsilon,\omega)$ and $\pm P_2(\epsilon,\omega).$
Here, two cases must be considered with regard to the multiplicity
 of these roots. 
One case is that in which the roots are all simple. In this case,
 the general solution of (\ref{phi2}) is
$
\Phi(s)=C_1e^{P_1s}+\tilde C_1e^{-P_1s}+C_2e^{P_2s}+\tilde C_2e^{-P_2s},
$
where $C_1$, $\tilde C_1$, $C_2$ and $\tilde C_2$ are complex constants.
The four BC given in (\ref{B.C.phi_head})
 and (\ref{B.C.phi_tail}) for $\Phi$ yield
%\begin{eqnarray*}
%&&\left(\begin{array}{cccc}
%1&1&1&1\\
%-P_2^2P_1&P_2^2P_1&-P_1^2P_2&P_1^2P_2\\
%P_1e^{P_1}&-P_1e^{P_1}&P_2e^{P_2}&-P_2e^{P_2}\\
%P_1^2e^{P_1}&P_1^2e^{P_1}&P_2^2e^{P_2}&P_2^2e^{P_2}\\
%\end{array}\right)
%\left(\begin{array}{c}
%C_1\\
%\tilde C_1\\
%C_2\\
%\tilde C_2
%\end{array}
%\right)\\
%&&
%\qquad\qquad\qquad
%\equiv \mbox{\boldmath $B$}\vec{C}=0.
%\end{eqnarray*}
\begin{eqnarray}
\left(\begin{array}{cccc}
1&1&1&1\\
-P_2^2P_1&P_2^2P_1&-P_1^2P_2&P_1^2P_2\\
P_1e^{P_1}&-P_1e^{P_1}&P_2e^{P_2}&-P_2e^{P_2}\\
P_1^2e^{P_1}&P_1^2e^{P_1}&P_2^2e^{P_2}&P_2^2e^{P_2}\\
\end{array}\right)
\left(\begin{array}{c}
C_1\\
\tilde C_1\\
C_2\\
\tilde C_2
\end{array}
\right)
%\equiv\mbox{\boldmath $B$}\vec{C}=0.\nonumber
=0.\nonumber
\end{eqnarray}
%We refer to the matrix {\boldmath$B$} as the BC matrix. The necessary
We refer to this matrix as the BC matrix {\boldmath$B$}. The necessary
% and sufficient condition for there to be a non-trivial $\vec C$ solving
 and sufficient condition for there to be a non-trivial $(C_1,\tilde C_1,C_2,\tilde C_2)$ solving
 this equation is
$\det${\boldmath$B$}$=0$. Because $\pm P_1$ and $\pm P_2$ are all complex
 functions of $\epsilon$
and $\omega$, $\det${\boldmath$B$}$=0$
 yields two equations for $\epsilon$ and
$\omega$. There exists one solution, which we found numerically to be
$(\epsilon_c,\omega_c)=(37.69...,191.25...)$ by searching 
in the region
$(\epsilon,\omega)\in[0,100]\oplus[0,1000]$.
The other case is that in which there is a multiple root of
$L(P)=0$.
 In this case,
the four roots are simply given by $\pm P_1=0$, 
$\pm P_2=\pm i\sqrt{\epsilon}\equiv\pm P_0$.
The general solution of (\ref{phi2}) is then
$
\Phi(s)=B_1+B_2s+B_3e^{P_0s}+\tilde B_3e^{-P_0s},
$
where $B_1,\,B_2,\,B_3$ and $\tilde B_3$ are complex constants. 
Again considering the
 BC matrix {\boldmath$B$}, we find that in this case,
 $\det${\boldmath$B$} is always nonzero. Thus, in this second case
there is no non-trivial solution that satisfies the BC in
 (\ref{B.C.phi_head}) and (\ref{B.C.phi_tail}).
Thus, in the case of a clamped tube, we have found 
only one solution in which the trivial solution might become
 unstable. This is the situation of all simple roots and occurs at
$\epsilon=\epsilon_c$, with a Hopf bifurcation of frequency $\omega_c$. 

Next, we consider the case of a hinged tube. We apply an  argument
 very similar to that above.
Now we consider (\ref{phi2}) with the four BC given by
 (\ref{B.C.phi_tail})
and
\begin{equation}
\label{B.C.phi_head_hi}
\Phi'(0)=\Phi'''(0)+\epsilon\Phi'(0)=0\quad\quad\mbox{hinged inlet}.
\end{equation}
Once again, we use the BC matrix method.
In this case, if we assume that the roots of $L(P)=0$ are all simple, 
 no solution is found numerically in the region
 $(\epsilon,\omega)\in[0,100]\oplus[0,1000]$.
 In the case that there is a multiple root of $L(P)=0$, the
 solution set is
$\{(\epsilon,\;\omega=0,\Phi (s)=B_0)\mid {^\forall}\epsilon >0,{^\forall
 B_0}\in \mbox{\boldmath $C$}\}$. This is the Goldstone mode,
 which exists due to the rotational
 symmetry of the hinged tube. This mode itself does
 not become unstable, but rather is marginally stable for all
 $\epsilon>0$. In
 the presence of the Goldstone mode, the null eigenspace may
 have a geometric multiplicity of 2. For this reason, we must seek
 a generalized null eigenfunction that satisfies
$
-\Phi''''-\epsilon\Phi''=B_0.
$
The general solution of this equation is
$
\Phi(s)=B_1+B_2s-\{B_0/(2\epsilon)\}s^2
       +B_3e^{P_0 s}+\tilde B_3e^{-P_0 s},
$
where $B_1,B_2,B_3$ and $\tilde B_3$ are complex constants.
The four BC given in (\ref{B.C.phi_tail}) and
 (\ref{B.C.phi_head_hi}) for $\Phi$ yield
\begin{displaymath}
\left(\matrix{
 0 & 1 & P_0 & -P_0 \cr
 0 & -P_0^2 & 0 & 0 \cr
 0 & 1 & P_0 e^{P_0} & -P_0 e^{-P_0}\cr
 0 & 0 & P_0^2 e^{P_0} & P_0^2 e^{-P_0}\cr
}\right)
\left(\matrix{ B_1 \cr B_2 \cr B_3\cr \tilde B_3\cr} \right) 
=\left(\matrix{ 0 \cr 0 \cr -B_0/P_0^2 \cr -B_0/P_0^2 \cr} \right).
\end{displaymath}
After some calculation, we find that the necessary and sufficient
 condition for $(B_1,B_2,B_3,\tilde B_3)\not=0$ is
 $\sqrt{\epsilon}=\tan{\sqrt{\epsilon}}$.
The smallest value of $\epsilon$ that satisfies this condition is
$\epsilon_h=20.16...$, and the null eigenfunction that
 is independent of the Goldstone mode is 
%\begin{displaymath}
%\Phi(s) = {B_0 \over 2} ({s^2 \over \epsilon_h} +
%          {\cos{\sqrt{\epsilon_h} s} \over 
%                           \epsilon_h^2\cos{\sqrt{\epsilon_h}}}).
%\end{displaymath}
$\Phi(s) = (B_0 /2 )\{s^2/\epsilon_h+(\cos{\sqrt{\epsilon_h} s})
/(\epsilon_h^2\cos{\sqrt{\epsilon_h}})\}$.
We therefore conclude that in the case of a hinged tube, it is possible that
 at $\epsilon=\epsilon_h$ the trivial solution
becomes unstable and that a pitchfork bifurcation involving a Goldstone
mode occurs. 
 
\begin{center} 
\begin{figure}[htbp]
 \epsfig{file=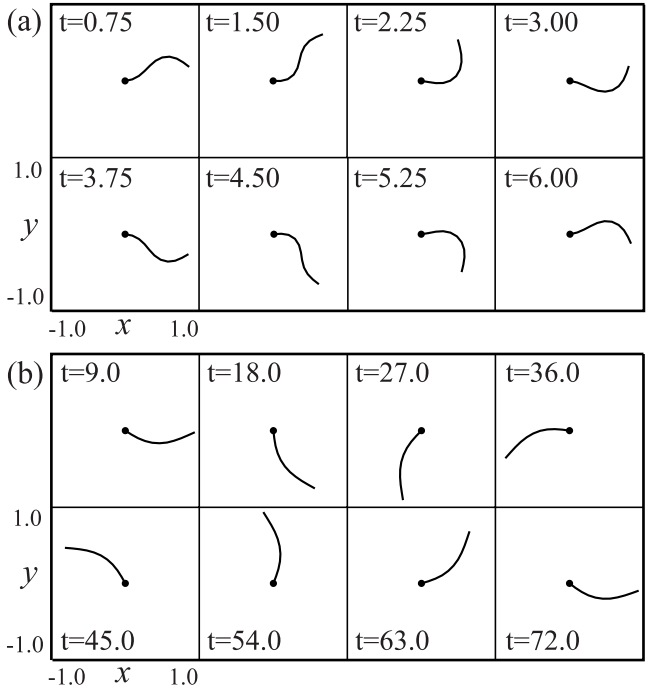}
 \caption{Numerical simulation of (1). (a)The clamped tube
undergoes a Hopf bifurcation and begins to flutter.
$N=10$, $\alpha =0.004$, $\beta=0.3$, $\gamma=1.0$.
(b) The hinged tube undergoes a pitchfork bifurcation involving a Goldstone
 mode and begins to rotate.
$N=10$, $\alpha =0.004$, $\beta=0.09$, $\gamma=1.0$.
 }
 \label{hopf}
\end{figure}
\end{center}

The above analysis has identified eigenmodes of the linearized system
that may be the first to become unstable.
The results suggest that the clamped tube undergoes a Hopf bifurcation
and begins to flutter, while
the hinged tube undergoes a pitchfork bifurcation involving a
Goldstone mode and begins to rotate.
 We confirmed this suggestion through the numerical
simulation of (\ref{discrete}).
%Here, we set $A=(L/N)^{-4}\alpha,B=(L/N)^{-2}\beta$, and $C=\gamma$.
Here, we set $A=l^{-4}\alpha,B=l^{-2}\beta$, and $C=\gamma$.
Figure \ref{hopf} displays some typical behavior
slightly above the first bifurcations, which are consistent with our
suggestion.

\begin{center}
\begin{figure}[htbp]
 \epsfig{file=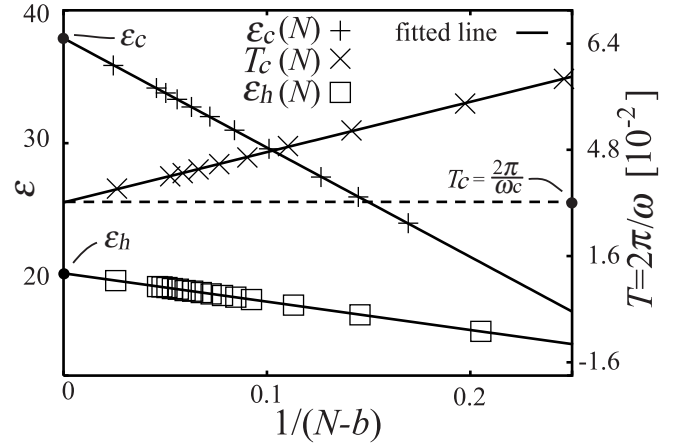}
 \caption{
The critical values of the discrete model,
 $\epsilon_c(N)$, $T_c(N)\equiv 2\pi/\omega_c(N)$, and $\epsilon_h(N)$,
 plotted as function of $1/(N-b)$,
  where $b=-0.9012$, $1.9353$, and $0.1126$, respectively, are obtained
 from the fit.
The intersections of the fitted lines and the $1/(N-b)=0$ axis 
correspond to the $N\to\infty$ limit. The uncertainty of the data is
 much smaller than the size of symbols in this plot.
 }
 \label{simulation}
\end{figure}
\end{center}

Figure \ref{simulation} shows that the critical values
 $\epsilon_c(N)$, $\omega_c(N)$, and $\epsilon_h(N)$ found numerically 
 converge to the theoretical values
 $\epsilon_c$, $\omega_c$, and $\epsilon_h$ as $N\to\infty$.
We calculated  $\epsilon_c(N)$, $T_c(N)\equiv 2\pi/\omega_c(N)$,
 and $\epsilon_h(N)$ with
 1\% accuracy for several values of $N$. 
 The data points can be fit to within the uncertainty on each data point
 by the function $a/(N-b)+c$, where $a$, $b$, and $c$ are the
 fitting parameters.
From this fitting we obtain the limiting values 
 $\epsilon_c(N)\to 38.0\pm0.9$, $\omega_c(N)\to 195.2\pm4.1$, and
 $\epsilon_h(N)\to 20.3\pm0.6$ as $N\to\infty$.
 These are almost identical to the theoretical values,
 37.69..., 191.25..., and 20.16....
Thus, we can say that the conjecture based on the results of linear
 stability analysis are
confirmed by the numerical simulation. Moreover, we confirmed
 numerically that both bifurcations are supercritical by examining 
 $\epsilon$ dependency of the magunitude of deformation.

Let us close this Letter with some supplementary comments.
%In the case of an inertial tube, there are two essential parameters
%If we should consider the inertial terms, there are two essential parameters
If the inertial terms should be considered, there are two essential parameters
in the equation of motion, rather than just one.
 This fact suggests that the behavior of an inertial tube may be much
 more complex.
Indeed, numerical simulations suggest
 that, while the inertial terms do not change
the qualitative properties of the first bifurcations,
they do change the subsequent bifurcations.
In particular,
the clamped inertial tube exhibits a period doubling
 route to chaos, while the hinged inertial
 tube undergoes a Hopf bifurcation as a second bifurcation and then
 exhibit a period doubling route to chaos.
Such bifurcations are not observed in the case of a non-inertial tube.

An interesting feature of our model is that is relevant to both
 aspirating and discharging tubes\cite{7}.
 In our model, the flow force is due
 only to the change in the momentum of the fluid at each point of the
 tube. The behavior exhibited by our model thus provides a prediction
 for both a discharging tube and an aspirating tube.

The authors are grateful to Y. Kuramoto, K. Sekimoto,
 S. Toh, G. Paquette and T. Miyoshi for informative discussions.
% now the references. delete or change fake bibitem. delete next three
%   lines and directly read in your .bbl file if you use bibtex.

% \begin{figure}
% \caption{}
% \label{}
% \end{figure}

% \begin{table}
% \caption{}
% \label{}
% \end{tabular}
% \end{table}

\end{document}